\documentclass[12pt]{article}
\pdfoutput=1

\usepackage{amsmath,amssymb,amsfonts,epsfig,cite,setspace,bigstrut,framed,eufrak}
\usepackage[all]{xy}
\usepackage{color}
\usepackage{pifont}

\makeatletter \@addtoreset{equation}{section} \makeatother
\renewcommand{\theequation}{\thesection.\arabic{equation}}

\begin{document}

\vskip 0.25in

\newcommand{\todo}[1]{{\bf\color{blue} !! #1 !!}\marginpar{\color{blue}$\Longleftarrow$}}
\newcommand{\nn}{\nonumber}
\newcommand{\comment}[1]{}
\newcommand\T{\rule{0pt}{2.6ex}}
\newcommand\B{\rule[-1.2ex]{0pt}{0pt}}

\newcommand{\CO}{{\cal O}}
\newcommand{\cI}{{\cal I}}
\newcommand{\cM}{{\cal M}}
\newcommand{\cW}{{\cal W}}
\newcommand{\cN}{{\cal N}}
\newcommand{\cR}{{\cal R}}
\newcommand{\cH}{{\cal H}}
\newcommand{\cK}{{\cal K}}
\newcommand{\cT}{{\cal T}}
\newcommand{\cZ}{{\cal Z}}
\newcommand{\cO}{{\cal O}}
\newcommand{\cQ}{{\cal Q}}
\newcommand{\cB}{{\cal B}}
\newcommand{\cC}{{\cal C}}
\newcommand{\cD}{{\cal D}}
\newcommand{\cE}{{\cal E}}
\newcommand{\cF}{{\cal F}}
\newcommand{\cA}{{\cal A}}
\newcommand{\cS}{{\cal S}}
\newcommand{\cX}{{\cal X}}
\newcommand{\IA}{\mathbb{A}}
\newcommand{\IP}{\mathbb{P}}
\newcommand{\IQ}{\mathbb{Q}}
\newcommand{\IH}{\mathbb{H}}
\newcommand{\IR}{\mathbb{R}}
\newcommand{\IC}{\mathbb{C}}
\newcommand{\IF}{\mathbb{F}}
\newcommand{\IS}{\mathbb{S}}
\newcommand{\IT}{\mathbb{T}}
\newcommand{\IV}{\mathbb{V}}
\newcommand{\II}{\mathbb{I}}
\newcommand{\IZ}{\mathbb{Z}}
\newcommand{\re}{{\rm Re}}
\newcommand{\im}{{\rm Im}}
\newcommand{\tr}{\mathop{\rm Tr}}
\newcommand{\ch}{{\rm ch}}
\newcommand{\rk}{{\rm rk}}
\newcommand{\ext}{{\rm Ext}}
\newcommand{\bi}{\begin{itemize}}
\newcommand{\ei}{\end{itemize}}
\newcommand{\beq}{\begin{equation}}
\newcommand{\eeq}{\end{equation}}
\newcommand{\bea}{\begin{eqnarray}}
\newcommand{\eea}{\end{eqnarray}}
\newcommand{\ba}{\begin{array}}
\newcommand{\ea}{\end{array}}

\newcommand{\CN}{{\cal N}}
\newcommand{\y}{{\mathbf y}}
\newcommand{\z}{{\mathbf z}}
\newcommand{\C}{\mathbb C}\newcommand{\R}{\mathbb R}
\newcommand{\CA}{\mathbb A}
\newcommand{\CP}{\mathbb P}
\newcommand{\cP}{\mathcal P}
\newcommand{\tmat}[1]{{\tiny \left(\begin{matrix} #1 \end{matrix}\right)}}
\newcommand{\mat}[1]{\left(\begin{matrix} #1 \end{matrix}\right)}
\newcommand{\diff}[2]{\frac{\partial #1}{\partial #2}}
\newcommand{\gen}[1]{\langle #1 \rangle}

\newtheorem{theorem}{\bf THEOREM}
\newtheorem{proposition}{\bf PROPOSITION}
\newtheorem{observation}{\bf OBSERVATION}

\def\theequation{\thesection.\arabic{equation}}
\newcommand{\setall}{
	\setcounter{equation}{0}
}
\renewcommand{\thefootnote}{\fnsymbol{footnote}}

\begin{titlepage}
\vfill
\begin{flushright}
{\tt\normalsize KIAS-P19054}\\

\end{flushright}
\vfill

\begin{center}
{\Large\bf  Topology of Generalized Spinors \\ \vskip 3mm and Chiral Anomaly}

\vskip 1.5cm

Ho-Ung Yee$^\dagger$ and Piljin Yi$^\ddagger$
\vskip 5mm
{\it$^\dagger$Department of Physics, University of Illinois,
Chicago, Illinois 60607, and\\
Kadanoff Center for Theoretical Physics, University of
Chicago, Illinois 60637, USA}\vskip 0.2cm
{\it$^\ddagger$School of Physics,
Korea Institute for Advanced Study, Seoul 02455, Korea}

\end{center}
\vfill

\begin{abstract}
Weyl fermions with nonlinear dispersion have appeared in
real world systems, such as in the Weyl semi-metals and
topological insulators. We consider the most general form
of Dirac operators, and study its topological properties
embedded in the chiral anomaly, in the index theorem, and
in the odd-dimensional partition function, by employing
the heat kernel. We find that all of these topological
quantities are enhanced by a winding number defined by
the Dirac operator in the momentum space, regardless of 
the spacetime dimensions. The chiral anomaly in $d=3+1$,
in particular, is also confirmed via the conventional
Feynman diagram. These interconnected results allow
us to clarify the relationship between the chiral anomaly
and the Chern number of the Berry connection, under dispute in
some recent literatures, and also lead to a compact proof
of the Nielsen-Ninomiya theorem.

\end{abstract}

\vfill
\end{titlepage}

\tableofcontents


\section{Introduction}

In quantum field theories with fermions, we often encounter topological
properties, with the chiral anomaly being perhaps the best known such effect.
Although initially derived from Feynman diagram \cite{Adler:1969gk,Bell:1969ts}, its topological nature became
quickly apparent via Fujikawa's alternative explanation \cite{Fujikawa:1979ay} as the failure of the
measure to be invariant under chiral rotations, which in turn translates to
the Atiyah-Singer index density of the Dirac operator.

In recent years, chiral anomalies and other topological aspects
emerged as relevant and useful concepts in condensed matter systems
as well, notably in Weyl semi-metals
and topological insulators \cite{Wan:2011udc,Burkov:2011ene,Xu:2011dn}.
In the study of such systems, one encounters fermion systems of more
general kind than those familiar to high energy physics. Instead of
the usual Dirac operator, linear in the spacetime derivatives,
a modified Hamiltonian of type
\bea
H\sim \sigma^+ (-iD_+)^n +\sigma^- (-iD_-)^n +\sigma^3(- iD_3) \ ,
\eea
on a flat spatial $\IR^3$ has appeared in the context of the Weyl semi-metal.
A chiral two-component fermion with such a generalized Dirac operator
as the Hamiltonian is expected to suffer  ``$n$" times the usual chiral
anomaly. This was initially motivated by merging of a pair of chiral
Dirac cones in the Brillouin zone, while more direct demonstrations
via the Fujikawa method were recently given for $n=2$ and $n=3$ \cite{Huang:2017rpa,Lepori:2018vwg}.
Although this Hamiltonian is natural from coalescence of
several Weyl cones in the Brillouin zone, its topological equivalence
to multiple Weyl fermions, as manifest in the anomaly, is hardly
immediate from the usual continuum field theory viewpoint.

This begs for general inquiries into the anomaly and other
topological aspects for fermions whose spacetime Dirac
operator takes the most general form
\bea
\gamma^\mu \cP_\mu (-iD) \ ,
\eea
where $-iD_\mu$ is the covariant momentum operator and $\cP_\mu$
polynomials, or even an arbitrary smooth functions thereof.
In the end, we will compute the index density, the anomaly, and
also the anomalous phase of the partition function
in odd spacetime dimensions, and find that  all of these are
minimally modified by the winding number of the
map $K_\mu \rightarrow\cP_\mu (K)$. Apart from this overall
factor, the structure of the anomaly and the phase of the
partition function remains intact.
The same computation can be manipulated to show that
this winding number is alternatively  computed by counting of
the critical points $K_*$, defined by $\cP(K_*)=0$, weighted by parities.

The latter should be reminiscent of the Morse theory \cite{Witten:1982im}
for those who are familiar with index theorems, but at the same time,
this alternative picture shows how the winding number information
of the generalized Dirac operator is connected to the Dirac/Weyl
cones in the momentum space, in a way that has been fruitfully
used in the condensed matter literatures. In particular, the two alternative
interpretations via the winding number and the Morse counting
represent, respectively, the ultraviolet and the infrared viewpoint
of one and the same quantity. The former viewpoint will connect
to topological objects known in the momentum space as the Berry
monopole, whose quantized flux can be also related to the $d=2+1$
topological insulator in the condensed matter literatures \cite{Thouless:1982zz,Chang:1996zz,Sundaram1999,Bliokh2005,Fu:2007uya,Moore:2006pjk,Qi:2008ew}.

Extending the discussion to odd spacetime dimensions, one finds
a similar modification of the anomalous phase of the partition
function. Given an odd-dimensional Dirac operator, this phase
is computed by the eta-invariant which in turn are related to
the Chern-Simons action. We will also see how this Chern-Simons
effective action is also multiplicatively enhanced by the same
kind of the winding number as in even dimensions.
If we consider this odd-dimensional spacetime as a flat boundary
of an even dimensional half space-time, an Atiyah-Patodi-Singer
index theorem holds, again with the new overall multiplicative
factor by the same winding number. This also means that the
connection via APS index theorem \cite{Witten:2015aba} between
the $d=2+1$ boundary fermions and a bulk $d=3+1$ topological
field theory carries over verbatim: much as in even dimensional
anomalies, the odd-dimensional anomalous phase does not
distinguish between $N$ ordinary Dirac fermions and a
generalized Dirac fermion with the winding number $N$.

It is our aim to derive these general results, and to explore
their physical consequences. The starting point of this investigation
is the chiral anomaly for such generalized Weyl fermions.
which is one of the most robust handles we have in all of quantum
fermions with continuous classical symmetries. The chiral anomaly,
after many decades of its initial discovery, can be still mysterious.
On the one hand, it is an infrared phenomenon of anomalous
particle creation and annihilation at zero energy (level crossing point),
in background field configurations where both parity and time-reversal
symmetries are broken. On the other hand, its topological nature makes
it computable also in ultraviolet scales, leading to its expression
in terms of local topological density of background fields.
This infrared-ultraviolet connection is a profound characteristic of
chiral anomaly, which, when formulated in Euclidean space, leads to
its deep connection to the index theorems in mathematics.

The infrared-ultraviolet connection of the chiral anomaly may also
manifest itself in momentum space. In the infrared view point,
the anomaly should be given by contributions from local level
crossing points, where in/out-flows of particle numbers happen.
Since the particle number is conserved away from these points
due to the Liouville theorem \cite{Xiao:2005qw}, the same anomaly may also be seen
in the ultraviolet region of large momenta, captured by some
topology of the theory in consideration.

Can we prove the existence of such an infrared-ultraviolet
connection in momentum space? If yes,
what topology of the fermion theory in large momenta contains
the information of the infrared chiral anomaly?  One of our
main results in this work is to provide a rigorous answer to
this question. We show the existence of an infrared-ultraviolet
connection of chiral anomaly in momentum space for a general
class of theories, where the Dirac operator is an arbitrary
polynomial of covariant derivatives. In particular, we prove in Section 3.1
that the topology of Berry's curvature of projected chiral
spinor in the asymptotically large momentum region precisely carries
the same information of chiral anomaly in the infrared.

We hope that our work answers some of the questions raised in
\cite{Fujikawa:2005cn,Fujikawa:2017ych,Mueller:2019gjj} regarding
the connection between chiral anomaly and
the Berry's curvature in momentum space.
The Berry's curvature of chiral spinors is an essential ingredient
of the kinetic description of chiral particles in phase space, the
chiral kinetic theory \cite{Son:2012wh,Stephanov:2012ki,Chen:2012ca,Duval:2014ppa,Mueller:2017arw}, where semi-classical approximation
is justified at large momenta. It is responsible for many novel
transport phenomena in real-time dynamics of (pseudo) chiral
fermion systems, in both condensed matter physics of Dirac/Weyl
semi-metals \cite{Nielsen:1983rb,Son:2012wh,Zyuzin2012,Goswami:2012db,Basar:2013iaa,Landsteiner:2013sja,Ma:2017loe} and the physics of quark-gluon plasma in relativistic
heavy-ion collisions \cite{Kharzeev:2007jp,Kharzeev:2013ffa}.

This includes most notably the Chiral
Magnetic Effect \cite{Fukushima:2008xe,Vilenkin:1980fu,Kharzeev:2007tn}, the Chiral Vortical Effect \cite{Erdmenger:2008rm,Banerjee:2008th},
and the Anomalous Hall Effect \cite{Jungwirth:2002zz,Fang2003,Yao2004,Haldane2004}.\footnote{
See \cite{Li:2014bha,Huang:2015eia,Zhang:2016ufu} for the experimental observation of Chiral
Magnetic Effect in Weyl semimetals, and see \cite{Kharzeev:2015znc,Skokov:2016yrj} for
the recent status of experimental search of Chiral Magnetic
Effect in relativistic heavy-ion experiments.}
Within the kinetic theory description, it has been argued that
chiral anomaly may also be explained by the same Berry's
curvature \cite{Son:2012wh,Stephanov:2012ki,Dwivedi:2013dea}. Since the kinetic theory description as well
as the concept of Berry's phase breaks down near level
crossing points where chiral anomaly happens, a more
rigorous treatment is needed to justify such a relation
between chiral anomaly and the Berry's phase in momentum
space \cite{Fujikawa:2005cn,Fujikawa:2017ych,Mueller:2019gjj}.
The infrared-ultraviolet connection we show in this work
fills the missing logical gap between the two.

For the most part of this note, we will employ the heat kernel
methods \cite{heat}, as it is universally applicable to all spacetime
dimensions and is very effective for extracting topological
information. In Section 4, however, we will also resort to
the usual triangular Feynman diagram for $d=4$ chiral anomaly,
where the modification of the current operators, on top of
the higher inverse power of the propagator, plays a crucial role.

\section{Generalized Spinors and Dirac Operators}

We would like to consider a Dirac index problem with the
operator generalized as
\bea\label{gDirac}
\cQ=\gamma^\mu\cP_\mu(-iD)
\eea
with smooth functions $\cP_\mu$. Let us take the Dirac
matrices in the chiral basis,
\bea
\gamma^a=\left(\begin{array}{cc} 0&\sigma^a \\ \sigma^a&0\end{array}\right)\ , \qquad
\gamma^4=\left(\begin{array}{cc} 0&-i\\ i&0\end{array}\right)\ ,
\eea
and we use the covariant derivative
\bea
D_\mu =\partial_\mu + A_\mu \ ,
\eea
with anti-hermitian gauge field $A_\mu$. The Dirac operator has the form
\bea
\cQ =  \left(\begin{array}{cc}0 & \cD \\ \cD^\dagger & 0\end{array}\right)
\eea
with
\bea
\cD =\left(\sigma^a\cP_a -i  \cP_4\right) \ , \qquad \cD^\dagger =\left(\sigma^a\cP_a +i \cP_4\right) \ .
\eea
We are interested in the index theorem of $\cQ$, and the chiral
anomaly associated with a Weyl fermion with the kinetic operator $\cD$.

\subsection{Index Density and Chiral Anomaly}

As is well known from the Fujikawa method \cite{Fujikawa:1979ay},
the failure of the chiral rotation of the path integral for the
relevant two-component Weyl fermion,
\bea
\int[D\bar\psi D\psi]\;e^{\int \bar\psi \cD\psi}
\eea
is measured by the index density, which can be written formally as
\bea\label{bulk}
{\rm Tr}\left(\Gamma\right)\equiv \lim_{s\rightarrow 0}{\rm Tr}\left(\Gamma e^{-s\cQ^2}\right)
\eea
with $\Gamma=-\gamma^1\gamma^2\gamma^3\gamma^4$. Note that the trace
here is over the 4-component Dirac spinors even though the physical
system is that of a Weyl spinor. One can understand this
from the well-known fact that, in the Euclidean
signature, $\bar\psi$ has to be treated
as independent and transforms oppositely to $\psi$ under the chiral
rotation. In practice, $\psi$ and $\bar\psi$ together define a
Dirac spinor, for which the formal index problem follows. For a most
comprehensive study of anomaly and the connection to the index
theorem, we refer readers to Ref.~\cite{AlvarezGaume:1983ig}.

Here, we proceed to compute this quantity by modifying the usual
heat kernel method, or
\bea
\lim_{s\rightarrow 0}{\rm Tr}\left(\Gamma e^{-s\cQ^2}\right) =
\lim_{s\rightarrow 0}\int d^4 x\;{\rm tr}\left(\Gamma G_s(x;x)\right)\ ,
\eea
where $G_s(y;x)\equiv \langle y\vert e^{-s\cQ^2}\vert x\rangle$ obeys
\bea
-\partial_s G_s(y;x) =\cQ^2 G_s(y;x)\ , \qquad
\lim_{s\rightarrow 0}G_s(y;x)=\delta^{(4)}(y-x)\ .
\eea
So, the problem boils down to how one computes $G_s(x;x)$.
For this, we start with
\bea\label{Q1}
\cQ^2=\cP_\mu\cP_\mu +\frac14[\gamma^\mu\,\gamma^\nu][\cP_\mu,\cP_\nu]\ ,
\eea
which we further split as
\bea\label{Q2}
\cQ^2=\cQ_0^2 +\delta\cQ^2 \ ,\qquad  Q_0^2\equiv  \cP_\mu(-i\partial)\cP_\mu(-i\partial)\ .
\eea
With the latter, we can perform the usual heat kernel expansion
\bea\label{heat}
G_s(y;x)& =&\sum_{l=0} G_s^{(l)}(y;x)\cr\cr
G_s^{(l+1)}(y;x)& =& -
\int_0^s dt \int d^4z\; G_{s-t}^{(0)}(y;z)\,\delta\cQ^2\, G_t^{(l)}(z;x) \ ,
\eea
where the free heat kernel
\bea
 G_s^{(0)}(z;x) = \langle z\vert e^{-s\cQ_0^2}\vert x\rangle \ ,
\eea
which is easily found
\bea\label{zero}
G_s^{(0)}(x+X;x) = \int\frac{d^4K}{(2\pi)^4} e^{iK\cdot X} e^{-s\cP(K)^2}
\eea
in the momentum space $\tilde \IR^4$ of $K_\mu$.

For the index density, the crucial step is the power counting of small $s$
in (\ref{heat}). Each iteration brings down a factor of $s$ given the $s$-integral,
but further fractional factors of $s$ arises from the $z$-integral combined
with operators in $\delta \cQ^2$. Note, in particular, that each derivative in
$\delta\cQ^2$ will cost some inverse fractional power of $s$. One key
identity will be
\bea\label{2nd}
\frac{1}{{\sqrt\pi}^4} \int d^4K\; {\rm det}\left(\frac{\partial \cP_\mu}{\partial K_\alpha}\right) e^{-s\cP(K)^2} = s^{-2}N_\cP \ ,
\eea
where $N_\cP$ is the asymptotic winding number of the map, $K\rightarrow \cP(K)$.
In other words, $N_\cP$ measures the multiplicity of the map over the target
$\IR^4$ with the orientation taken into account.
With (\ref{Q1}) and (\ref{Q2}), and with the insertion of $\Gamma$ in
(\ref{bulk}), it is clear that the first nontrivial expression out of (\ref{heat})
will occur at the second iteration, where we expect to find something like (\ref{2nd}) times $s^2$,
leading us to $N_\cP$ in the end. Let us now track how this occurs.

The relevant contribution can be found from the further expansion of
the squared Dirac operator
\bea \label{comm}
\frac14[\gamma^\mu ,\gamma^\nu][\cP_\mu,\cP_\nu]=\frac12\gamma^\mu\gamma^\nu
F_{\alpha \beta}\frac{\partial \cP_\mu(K)}{\partial K_\alpha}
\frac{\partial \cP_\nu(K)}{\partial K_\beta}+\cdots \ ,
\eea
where the ellipsis denotes terms that come with less free standing
derivatives, i.e., less factors of $K$'s, or more $A$'s. These cost less
power of $s^{-1}$ and effectively disappear as $s\rightarrow 0$ limit is
taken in the end. Because of the $\Gamma$ insertion in (\ref{bulk}),
the first nontrivial term arises in the second order of the iteration of
(\ref{heat}), when one pulls down $\gamma\gamma F$ in (\ref{comm}). This
will be accompanied effectively by a factor of $s^2/2$, due to the two
$s$ integrals, producing a term like (\ref{2nd}). This shows how all
the subsequent terms in (\ref{comm}) become irrelevant for the purpose
of computing the index density.
In fact, all interaction pieces in $\cP(-iD)^2$ belong to
the latter category, so for the purpose of computing the index density,
all that matter is the first term on the right hand side of (\ref{comm})
in place of $\delta \cQ^2$.

Let us trace this process more explicitly. Since an explicit factor of
$x$'s in $\delta \cQ^2$, such as in Taylor expansion of $F$'s, cost positive
factors $s$'s, relative to the one in (\ref{comm}), we only need to worry
about how the free-standing derivatives in $\delta \cQ^2$ works in the
heat kernel expansion. With
\bea
\Pi(K,;F)\equiv \frac12\gamma^\mu\gamma^\nu
F_{\alpha \beta}\frac{\partial \cP_\mu(K)}{\partial K_\alpha}
\frac{\partial \cP_\nu(K)}{\partial K_\beta}\ ,\qquad
\hat\Pi(-i\partial\,;F)\equiv \Pi(K;F)\biggr\vert_{K\rightarrow -i\partial} \ ,
\eea
one finds\footnote{Here we used
\bea
\int {d^4Y} e^{i(W-K)\cdot Y}=(2\pi)^4\delta^{(4)}(W-K)\ .
\eea
}
\bea
G^{(1)}_s(x+X;x) &=&\int_0^s dt \int d^4Y \;G_{s-t}^{(0)}(x+X;x+Y)\hat\Pi(-i\partial;F) G_t^{(0)}(x+Y;x)+\cdots\cr\cr
&=& \int_0^s dt \int d^4Y \int\frac{d^4K}{(2\pi)^4}\; e^{iK\cdot(X-Y)} e^{-(s-t)\cP(K)^2}\cr\cr
&&\hskip 3cm \times \Pi(W;F) \int\frac{d^4W}{(2\pi)^4}\; e^{iW\cdot Y} e^{-t\cP(W)^2}+\cdots\cr\cr
&=& s \int\frac{d^4K}{(2\pi)^4}\; e^{iK\cdot X} e^{-s\cP(K)^2}\times \Pi(K;F(x)) +\cdots \ .
\eea
The above is from an expansion of
the operator $\cQ^2$ around a generic point $x$, and the momentum
$K$ is conjugate to the ``small" displacement $X$.

It is clear that the momentum factors pile up through the iteration,
and since the position-dependence of $F_{\mu\nu}(x)$ does not enter
in the small $s$ limit, the iteration can be performed in the
momentum space straightforwardly.
Repeating one more time, the same computation gives
\bea
G^{(2)}_s(x+X;x)
&=&\frac{s^2}{2} \int\frac{d^4K}{(2\pi)^4}\; e^{iK\cdot X} e^{-s\cP(K)^2}\times (\Pi(K;F(x)))^2+\cdots \ .
\eea
Now we are ready to compute the index density:
\bea
&&\lim_{s\rightarrow 0}{\rm Tr}\left(\Gamma e^{-s\cQ^2}\right) = \lim_{s\rightarrow 0}\int d^4 x\;{\rm tr}\left(\Gamma G_s(x;x)\right)\cr\cr
&=&\lim_{s\rightarrow 0}\;\frac{s^2}{2} \int d^4 x\; {\rm tr}\left(\frac14\Gamma \gamma^\mu\gamma^\nu\gamma^{\mu'}\gamma^{\nu'}\right)F_{\alpha\beta}F_{\alpha'\beta'}\cr\cr
&&\hskip 1cm\times \int \frac{d^4K}{(2\pi)^4}\frac{\partial \cP_\mu(K)}{\partial K_\alpha}
\frac{\partial \cP_\nu(K)}{\partial K_\beta}\frac{\partial \cP_{\mu'}(K)}{\partial K_{\alpha'}}
\frac{\partial \cP_{\nu'}(K)}{\partial K_{\beta'}}e^{-s\cP(K)^2} \ .
\eea
With $\Gamma=-\gamma^1\gamma^2\gamma^3\gamma^4$ inserted,
we need to collect four distinct $\gamma$'s to ensure nonzero result,
and then the fermionic trace above gives
$-4\epsilon^{\mu\nu\mu'\nu'}$. Then, we may invoke (\ref{2nd}) and find
\bea\label{Anomaly}
\lim_{s\rightarrow 0}{\rm Tr}\left(\Gamma e^{-s\cQ^2}\right)
=-\frac{N_\cP}{32\pi^2} \int d^4x\;\epsilon^{\alpha\beta \alpha'\beta'}F_{\alpha\beta}F_{\alpha'\beta'}
=N_\cP\cdot \left(-\frac{1}{8\pi^2} \int F \wedge F\right) \ ,
\eea
which shows the usual index density, and hence the chiral anomaly is
enhanced by a factor of $N_\cP$, the winding number associated with
the map $K\rightarrow \cP(K)$.

Although we have computed the index density in four dimensions, the
generalization to arbitrary even dimensions, $d$, is immediate,
and gives
\bea
\lim_{s\rightarrow 0}{\rm Tr}\left(\Gamma e^{-s\cQ^2}\right)
=N_\cP\cdot \left(\frac{1}{(d/2)!(2\pi i)^{d/2}} \int F\wedge\cdots \wedge F\right)
\eea
with $d/2$ number of  the field strength 2-form $F=F_{\mu\nu}dx^\mu\wedge dx^\nu/2$,
anti-hermitian as before. The winding number of the map $K\rightarrow \cP(K)$
enters this formula via\footnote{We introduced
$\tilde\partial$ to emphasize that it is a partial derivative
in the momentum space.}
\bea\label{NcPd}
N_\cP\equiv \frac{1}{{\sqrt\pi}^d} \int_{\IR^d} d^d K\; {\rm det}\left(\tilde\partial^\alpha \cP_\mu\right) e^{-\cP(K)^2}\ ,
\eea
Each oriented copy of $\hat\IR^d$ in the image gives 1 times the
sign of the Jacobian, so this measures how many
times the map $\cP(K)$ covers the target $\hat\IR^d$.

\subsection{Infrared Interpretation}

Note that this integral produces an integer, as long as $\cP^2$ is asymptotically unbounded,
since
\bea
N_\cP =\frac{1}{{\sqrt\pi}^d} \int d^d  \cP\;e^{-\cP^2}\ ,
\eea
where $N_\cP$ is now hidden in the integration domain; The integral
on the right hand side gives 1 for each integration domain of
$\hat\IR^d$, but this is multiplied by $N_\cP$ since the map
$K\rightarrow \cP$ is $N_\cP$-fold cover of $\hat \IR^d$. This topological
characterization may be considered an ultraviolet description since the winding
number is defined via the asymptotic behavior of the map $\cP(K)$.

On the other hand, the index and the anomaly are fundamentally
infrared phenomena, so should be equally visible in the small
$|\cP|$ limit. For this, note that the expression is invariant under
$\cP\rightarrow C\cdot \cP$
for any positive real number $C$, which we already used to scale away
$s$ above to reach (\ref{NcPd}). Going back to $K$-space integral and
taking a limit of $C\rightarrow \infty$, however, we see that (\ref{NcPd})
localizes at the critical points, $\cP=0$, and the winding number has
an alternative form, as a sum over the critical points, weighted by $\pm 1$,
depending on the sign of the determinant there,
\bea
\sum_{\{K_*\vert \cP(K_*)=0\}}1\cdot{\rm sgn}
\left[{\rm det}(\tilde\partial^\alpha \cP_\mu)\right]\biggr\vert_{ K= K_*}\ ,
\eea
provided that all the critical points are non-degenerate, i.e., provided that
the determinants there do not vanish. In fact, it is easy to see how this
generalizes to a case with degenerate critical points,
\bea
\sum_{\{K_*\vert \cP(K_*)=0\}}N_\cP(K_*)\ ,
\eea
where $N_\cP(K_*)$ is the local winding number near such (degenerate)
critical points. This is a Morse theory counting if $\cP_\mu=\tilde\partial_\mu\, W(K)$
for some Morse function $W(K)$ \cite{Witten:1982im}, although we do not
really need the latter here.

Although we started with the expression
(\ref{NcPd}) that has a natural interpretation as a winding number,  measured
at the asymptotic region of $K$-space, this alternative description counts the
critical points, $\cP=0$, where the Dirac operator $\cQ$ may be approximated
by a linear form
$$
\cQ\simeq \tilde\partial^\alpha \cP_\mu(K_*) \,(-i\gamma^\mu D_\alpha)\ .
$$
One merely counts how many approximate Dirac cones appear in the
infrared end of the dynamics, whose chiralities are dictated by the matrix
$\tilde\partial^\alpha \cP_\mu(K_*)$.
If the latter has negative determinant, this can be translated to a chirality
flip, relative to others with positive determinant.

\section{Non-Relativistic Iso-Spinor}\label{sec:NR}

For condensed matter systems, one sometimes
encounters such generalized Weyl fermions where the
two components  actually refer to
flavors or ``isospin" rather than the real spin
associated with the angular momentum. In recent years,
the chiral anomaly in this isospin context has surfaced
as important issues, so let us apply what we have
developed in the previous section to these real
systems.

Since these are all non-relativistic fermions,
the direction 4 plays  a special role as the
genuine (Euclidean) time direction and as such
we will be content with the single derivative
there. As such, we may specialize to the case
\bea\label{nonR}
\cP_4=-iD_4 +\Delta(-i\vec D) \ ,\qquad \cP_a=P_a(-i\vec D)\ ,
\eea
where we split the four vector to the Euclidean time
component and a 3-vector distinguished by the arrow.
This will correspond to, in Lorentzian time, a two-component
Weyl Hamiltonian of type
\bea
H=\sigma^a P_a-\Delta \ .
\eea
The necessary Wick rotation prescription will become clearer
when we compute the anomaly by a Feynman diagram in next
section, but for now we will stick to this Euclidean viewpoint.

We could simply rely on the results of the previous section,
whereby the anomaly can be seen not affected by the presence
of $\Delta$ at all. Still, let us retrace part of these
steps for an illustration, with a simplifying assumption of $\Delta=0$.
In other words, let us consider
\bea\label{nonR4}
\cD=\sigma^a P_a( -i\vec D) - D_4 \ ,\qquad \cD^\dagger =\sigma^a P_a( -i\vec D) + D_4 \ ,
\eea
for some smooth functions $P_a$ of $-iD_{a=1,2,3}$.
In addition, we will assume that $ |\vec P(\,\vec k\,)|^2$
grows indefinitely at large $\vec k$'s;
this would be the case, for example, if $P$'s are generic polynomials.
Our experience above suggests that the anomaly must be again dictated by
a topology of the map
$
k_a\;\;\rightarrow\;\; P_a(\,\vec k\, )
$.

The zero-th order heat kernel is
\bea
G_s^{(0)}(x+X;x)& =&\frac{1}{\sqrt{4\pi s}} e^{-X_4^2/4s}\times
\int\frac{d^3k}{(2\pi)^3} e^{i\vec k\cdot \vec X } e^{-s\vec P^2}\ ,
\eea
where we now separate out the 3-vector $\vec X$ from 4-vector $X$, etc.
The pieces in $\cQ^2$ that can contribute to the index density are, as before
\bea
\frac14[\gamma^\mu,\gamma^\nu] [\cP_\mu, \cP_\nu] &=&
\sum_{a=1}^3\gamma^4\gamma^a F_{4h}\;\tilde\partial^h P_a(\,\vec k\,)+\cdots\cr\cr
&+&\sum_{b>c}\gamma^b\gamma^c F_{fg}\;\tilde\partial^f P_b(\,\vec k\,)\tilde\partial^g P_c(\,\vec k\,)+\cdots
\eea
expressed in the momentum space, and $F$ is considered slowly varying.
The structure of the index density we have seen implies that
the integrand will contain the multiplicative factor
\bea
\frac{1}{6}\epsilon^{abc}\epsilon_{hfg}(\tilde\partial^hP_a)( \tilde\partial^f P_b)(\tilde\partial^gP_c)={\rm det}\left(\frac{\partial P_a(\, \vec k\,)}{\partial k_h}\right)\ ,
\eea
and we can deduce the expression that plays the role of $N_\cP$
in (\ref{Anomaly}),
\bea\label{NvP}
N_{\vec P}\;\;\equiv\;\;  \frac{ s^{3/2}}{\sqrt{\pi}^3}\int d^3k\;{\rm det}\left(\frac{\partial P_a(\, \vec k\,)}{\partial k_h}\right)e^{-s\vec P^2}
\;\;=\;\; \frac{1}{\sqrt{\pi}^3}\int d^3P\;e^{-\vec P^2}\ ,
\eea
where the factor $s$ on the left hand side is scaled away without affecting the
result by the change of the integration variable $s^{1/2}\vec P\rightarrow \vec P$ on the
right hand side.

Again, the map $ \vec k\rightarrow  \vec P(\,\vec k\,)$ is
in general a multiple cover of the target $\hat\IR^3$, which translates
to the domain of the integral on the right hand side being several
copies of $\hat\IR^3$: The integral measures precisely this
multiplicity. Alternatively, this can be viewed as the winding number
of the map $ \vec k/|\vec k|\rightarrow \vec P(\,\vec k\,)/|P(\,\vec k\,)|$,
from $\tilde \IS^2$ to $\hat \IS^2$, in the large $|\vec k|$ limit. We
conclude that for the most general non-relativistic Weyl fermion in
3+1 dimensions, (\ref{NvP}) is the right coefficient to  the chiral anomaly.

\subsection{The Berry Monopoles in the Momentum Space}

We saw that the chiral anomaly of a general two-component
isospinor is given by the winding number of the map $\vec k\to \vec P(\,\vec k\,)$
in momentum space. We will show that the same topology
underlies the Berry's connection of projected chiral spinor
in momentum space.\footnote{What we outline here is  a well-known
computation in the context of $d=2+1$ topological insulators
generally. See for example Ref.~\cite{Qi:2008ew}. One way to realize the latter is by
imagining a two-dimensional Brillouin zone as a slice
in $d=3+1$ Brillouin zone of Weyl semi-metal,
between a pair of chiral Weyl and anti-chiral Weyl points.}
For this, one considers a two-level problem with the Hamiltonian
\bea
\sigma^a P_a(\,\vec k\,)\ ,
\eea
whose two eigenvalues are $\pm|\vec P(\,\vec k\,)|$. Denoting the two respective
eigenvectors by $\vert \pm\rangle$,  the Berry connection is
\bea
\cA^\pm = -\langle \pm\vert \frac{\partial}{\partial k_a}\vert \pm\rangle\, dk_a
\eea
A well-known result is that, when $P_a=k_a$, the Berry connection carries
the unit magnetic flux. Let us recapitulate this simpler case first. In the
spherical coordinates,
\bea
\sigma^a k_a =|\vec k|\left(\begin{array}{cc} \cos\theta & \sin\theta e^{-i\phi} \\\sin\theta e^{i\phi} &-\cos\theta\end{array}\right)\ ,
\eea
the two eigenvectors are
\bea
\vert + \rangle_{\sigma^a k_a} = \left(\begin{array}{c} \cos(\theta/2)e^{-i\phi/2}
\\ \sin(\theta/2) e^{i\phi/2}\end{array}\right)\ , \qquad
\vert - \rangle_{\sigma^a k_a} = \left(\begin{array}{c} -\sin(\theta/2)e^{-i\phi/2}
\\ \cos(\theta/2) e^{i\phi/2}\end{array}\right)\ .
\eea
Therefore, the two Berry connections are common up to a sign,
\bea
{\cal A}^\pm \biggr\vert_{\sigma^a k_a  }= \pm \frac{i}{2}\cos\theta d\phi \ ,
\eea
which carries a unit $2\pi $ magnetic flux over $\tilde \IS^2$, or the unit Chern
number. As usual this is up to
$U(1)$ gauge transformations under the innocuous phase rotations,
\bea
\vert\pm \rangle \quad\rightarrow \quad e^{i\Lambda_\pm(\vec k)}\vert \pm\rangle\ .
\eea

Coming back to $H=\sigma^aP_a$ is merely a matter of replacing
$\vec k$ by $\vec P(\,\vec k\,)$, so we can express
\bea\label{AP}
\cA^\pm \biggr\vert_{\sigma^a P_a }= \pm \frac{i}{2}\cos\Theta\, d\Phi
\eea
with the spherical angles $\Theta$ and $\Phi$ of $\vec P$, which needs to be pulled back
to $\vec k$ space. In other words, denoting the standard $2\pi$ Wu-Yang monopole field \cite{Wu:1976ge}
in $\vec P$ space by ${\bf A}$ and $\bf F$ its field strength, the Berry connection
and the field strength thereof in the momentum space are
\bea\label{BM}
\cA^\pm  =\pm {\bf A}_a\;\frac{\partial P^a}{\partial k_f}\;dk_f\ , \qquad
\cF^\pm  =\pm \frac12\,{\bf F}_{ab}\;\frac{\partial P^a}{\partial k_f}\frac{\partial P^b}{\partial k_g}\;dk_f\wedge dk_g \ .
\eea
Although (\ref{AP}) seemingly gives a unit $2\pi $ flux, this is not
generally the case because the map $\vec k\rightarrow \vec P$ can
be a multi-cover of the  target $\hat\IR^3$.

The actual integral that computes the Chern number is, with $P^*$ being
the pull-back of the map $\vec P(\,\vec k\,)$
\bea
\frac{1}{2\pi i} \int_{S_2} \cF^\pm  =\pm\frac{1}{2\pi}  \int_{S_2}P^*({\bf F}) =\pm \frac{1}{2\pi} \int_{B_3}P^*(d{\bf F})\ ,
\eea
where $S_2=\partial B_3$ is an arbitrary 2-surface in the
$\vec k$ space. In particular, if one takes $B_3=\tilde\IR^3$ the entire
momentum space,
\bea
\frac{1}{2\pi i} \int_{\tilde\IR^3}P^*(d{\bf F}) =\frac{1}{2\pi i} \oint_{\tilde \IS^2_\infty}P^*({\bf F})=N_{\vec P}\ ,
\eea
since at the asymptotic two-sphere $\tilde \IS^2_\infty$
the total flux is multiplied by the net winding number $N_{\vec P}$.

On the other hand, $d{\bf F}=0$ everywhere except at the
origin $\vec P=0$, so we may rewrite this integral as
\bea
\frac{1}{2\pi} \sum_{\{\vec k_*\vert \vec P(\,\vec k_*)=0\}} \int_{B^3_\epsilon(\,\vec k_*)}P^*(d{\bf F})\ ,
\eea
where $B_\epsilon^3(\,\vec k_*)$ is an infinitesimal 3-ball centered at the
$\vec k_*$. This in turn becomes,
\bea
\frac{1}{2\pi i} \sum_{\{\vec k_*\vert \vec P(\,\vec k_*)=0\}} \int_{\partial B^3_\epsilon(\,\vec k_*)}P^*({\bf F})\ ,
\eea
and, since $\bf F$ is a unit monopole in $\vec P$ space, we find
\bea\label{Nsum}
\sum_{\{\vec k_*\vert \vec P(\,\vec k_*)=0\}} N_{\vec P}(\,\vec k_*) \;\;=\;\; N_{\vec P}\ ,
\eea
where $N_{\vec P}(\,\vec k_*)$ is the winding number of $\vec P$, measured by
the infinitesimal neighborhood around $\vec k_*$, which of course sum up to $N_{\vec P}$.

The total flux is computed by taking $S_2$ equal to the
asymptotic $\tilde\IS^2$ of $\vec k$ space $\tilde\IR^3$, which maps to $N_{\vec P}$
times the asymptotic $\hat\IS^2$ of the target $\hat\IR^3$
if the map $\vec P$ covers the target $\hat\IR^3$ $N_{\vec P}$ times.
As such, this Berry connection carries precisely $2\pi N_{\vec P}$ total flux, where $N_{\vec P}$
may be also computed as in (\ref{NvP}), provided that $\vec P^2$ is
divergent everywhere asymptotically. This shows that
the chiral anomaly is precisely given by the same winding number
$N_{\vec P}$, offering a rigorous logical link between the
chiral anomaly and the Berry's connection in momentum space.
As should be clear from how we arrive at this connection via our
generalized index theorem, we emphasize that this is a non-trivial,
yet understandable manifestation of infrared-ultraviolet connection,
which is a fundamental characteristic of chiral anomaly.

\subsection{Infrared View and Real Material}

As we have already seen in Section 2, one can compute the same winding number
alternatively by rescaling $P_a\rightarrow C\cdot P_a$,
and going back to $d^3k$ integral. This  localizes
to small neighborhoods around $\vec k=\vec k_*$ where $\vec P(\,\vec k_*)=0$, and as
such really counts inverse images of $\vec P=0$ with weights $\pm 1$,
\bea\label{Morse3}
\sum_{\{\vec k_*\vert \vec P(\,\vec k_*)=0\}}1\cdot{\rm sgn}\left[{\rm det}(\tilde\partial^h P_a)\right]\biggr\vert_{\vec k=\vec k_*}\ .
\eea
This alternative form is viable when the critical points are
isolated and non-degenerate, where one is counting
Weyl cones with $\vec P$ locally linear $\vec k$.
If the sign happens to be negative, it has the same effect
as flipping signs of odd number of $\gamma^a$'s, resulting
in anti-Weyl fermions instead of Weyl fermions.
When we allow degenerate critical points, we find
\bea
\sum_{\{\vec k_*\vert \vec P(\,\vec k_*)=0\}} N_{\vec P}(\,\vec k_*)\ ,
\eea
the same as (\ref{Nsum}).

In real material the momentum $\vec k$ lives in a compact
Brillouin zone, $\tilde \IR^3/\Lambda$, where $\Lambda$
is the dual lattice to the crystalline lattice of the
material. On the flip side, $\vec P$ does not
extend indefinitely into $\hat\IR^3$ either. Instead,
$\vec P(\,\vec k\, )=\vec P(\,\Lambda\vec k\,)$.
One immediate question is how the story so far is
affected by such a compact Brillouin zone.
Consider a pair of energy bands that meets at one or more Bloch
momentum $\vec k=\vec k_*$, which for our purpose means a continuous map
from $\vec k$ in the Brillouin zone $\tilde\IR^3/\Lambda$ to a $2\times 2$
Hamiltonian $H(\,\vec k\,)$ such that
\bea
H(\,\vec k\,)=\vec \sigma\cdot\vec P(\,\vec k\,) - \Delta(\,\vec k\,)\ , \qquad \vec P(\,\vec k_*)=0\ .
\eea
The appearance $\Delta$ is of course necessary for real material,
but its connection to the same symbol in the above Euclidean
computation might look a bit unclear, since the naive Wick rotation
would render $\Delta$ in (\ref{nonR}) to become pure imaginary.
The answer to this is that one really starts with Lorentzian
signature and rotates the contour of $k_0$ such that one reaches
(\ref{nonR}) in the end. More on this would be elaborated in the
next section.

In both the Euclidean anomaly computation and the Berry phase
computation, we have seen that only the winding number
associated with $\vec P(\,\vec k\,)$ matters for these topological
characterization. In the target, one starts with a monopole of
Berry connection near $\vec P=0$, which we must pull-back
to the Brillouin zone $\tilde\IR^3/\Lambda $.
One puzzling aspect is that now $\vec P^2$ must be bounded
as it is defined on a compact Brillouin zone, to begin with,
and in real material the energy eigenvalues $-\Delta\pm|\vec P|$
must be bounded above and below: the integral formulae
such as (\ref{NvP}) appear to have no reason to produce an integer.
This quandary is saved by the observation that under such
circumstances (\ref{NvP}) always produces zero, as we see
below.

Let us replace
$\tilde\IR^3$ by $\tilde\IR^3/\Lambda$ and consider a surface $S_2 =\partial B_3$
that encloses all the critical points $P(\,\vec k_*)=0$, whereby we have
\bea
N_{\vec P}=\frac{1}{2\pi i} \int_{B_3} dP^*({\bf F})=\frac{1}{2\pi i} \int_{S_2} P^*({\bf F})\ .
\eea
On the other hand, since $\tilde\IR^3/\Lambda$ is closed, $-S_2$
is also a boundary to the complement $B_3^c$. Given that no
magnetic monopole exists in $B_3^c$, we have
\bea
N_{\vec P}=-\frac{1}{2\pi i} \int_{-S_2} P^*({\bf F}) = - \frac{1}{2\pi i} \int_{ B_3^c} dP^*({\bf F})=0\ ,
\eea
and further may as well shrink $B_3^c$
to nothing, so that $B_3=\tilde\IR^3/\Lambda$, and find
\bea
N_{\vec P}= \frac{1}{2\pi}\int_{\tilde\IR^3/\Lambda } d\left(P^*({\bf F})\right) =0\ .
\eea
In view of our infrared alternative, the same can be expressed as
\bea
\sum_{\{\vec k_*\vert \vec P(\,\vec k_*)=0\}} N_{\vec P}(\,\vec k_*) =0\ ,
\eea
or
\bea
\sum_{\{\vec k_*\vert \vec P(\,\vec k_*)=0\}}1\cdot{\rm sgn}\left[{\rm det}(\tilde\partial^h P_a)\right]\biggr\vert_{\vec k=\vec k_*} =0\ ,
\eea
if all the critical points are non-degenerate.

In the end, the number of Weyl points and the number of
anti-Weyl points are always equal, for any system on
a real space lattice, provided that $\vec P$ is smooth
enough in such a compact Brillouin zone. This is,
of course, nothing but the Nielsen-Ninomiya theorem \cite{Nielsen:1981hk}.

\subsection{Multiple and Degenerate  Weyl Semi-Metal}

Problems of this kind has been dealt with in recent literature
for a simple power-like $\vec P$.
The main prototype  is
\bea
\cD=\left(\sigma^+ (-iD_+)^2 +\sigma^- (-iD_-)^2 +\sigma^3(- iD_3) - D_4\right)\ ,
\eea
where $D_\pm= D_1\mp i D_2$ and $\sigma^\pm=(\sigma^1\pm i \sigma_2)/2$.
This has been motivated by a merging of a pair of Dirac
cones, initially separated in the Brillouin zone.
A slight generalization of this can be achieved by elevating
the power to an arbitrary positive $n$, and also replacing
$-iD_3$ by
\bea
P_3=C_l (-iD_3)^l+\cdots\ ,
\eea
where the ellipsis denotes lower order monomials of $-iD_3$.
As is clear from the general formalism, these subleading pieces
are irrelevant for the problems at hand, so we may as well consider
\bea
\cD=\left(\sigma^+ (-iD_+)^n +\sigma^- (-iD_-)^n +\sigma^3C_l(- iD_3)^l - D_4\right)\ ,
\eea
whose associated Dirac operator is
\bea
\cQ =  \left(\begin{array}{cc}0 & \cD \\ \cD^\dagger & 0\end{array}\right)=
\left(\gamma^+ (-iD_+)^n +\gamma^- (-iD_-)^n +\gamma^3 C_l(-i D_3)^l +\gamma^4 (-i D_4)\right)
\eea
with $\gamma^\pm=(\gamma^1\pm i \gamma^2)/2$. The anomaly computation for $n=2,3$ and $l=1$ with $C_1=1$ has
been performed in recent literatures \cite{Huang:2017rpa,Lepori:2018vwg}.
Although this class of examples clearly fall under the general
formalism above, we repeat the exercise in part as an
concrete example of our general formulation but also
to show explicitly how the additional power $l$ enters the story.

The zero-th order part of the squared Dirac operator is
\bea
-\cQ^2_0 =((\partial_1)^2+(\partial_2)^2)^n+C_l^2(\partial_3)^{2l}+(\partial_4)^2\ ,
\eea
so that, now with $X=(Z;U,V)$ and $K=(p;q,\tilde q)$
\bea
G_s^{(0)}(x+X;x)& =&\frac{1}{\sqrt{4\pi s}} e^{-V^2/4s}\times
\int\frac{d^2 p\,dq}{(2\pi)^3} e^{i(p\cdot Z+q U) } e^{-s (p_1^2+p_2^2)^n-sC_l^2q^{2l}}\ .
\eea
Again the key quantity to compute is the second pieces in
\bea
\cQ^2= \cP_\mu \cP_\mu+\frac14[\gamma^\mu,\gamma^\nu] [\cP_\mu, \cP_\nu]\ .
\eea
and the relevant terms in $\delta\cQ^2$  are
\bea
\frac14[\gamma^\mu,\gamma^\nu] [\cP_\mu, \cP_\nu] &=&
\gamma^1\gamma^2(n^2F_{12})(p_+p_-)^{n-1}+\cdots\cr\cr
&+&\gamma^\pm\gamma^3(-nl (F_{13}\mp iF_{23}))C_l q^{l-1}(p_+)^{n-1}+\cdots\cr\cr
&+&\gamma^\pm\gamma^4(-n (F_{14}\mp iF_{24}))(p_+)^{n-1}+\cdots\cr\cr
&+&\gamma^3\gamma^4 F_{34}\,lC_lq^{l-1}+\cdots\ .
\eea
As before, the momentum accumulates through each iteration,
and the relevant part of the heat kernel can be found at
the second order of iteration,
\bea
G_s(x+X;x)&=&\cdots\; -\Gamma\frac{s^{3/2}}{\sqrt{4\pi}}\,\left(l\times f(F(x)\right) e^{-V^2/4s}\cr\cr
&&\times \int \frac{d^2p \,dq}{(2\pi)^3} (p_+p_-)^{n-1}q^{l-1}
e^{ip\cdot Z+iqU} e^{-s((p^2)^n+C_l^2(q^2)^l)}+\cdots\ ,
\eea
with
\bea
f(F)= n^2(F_{12}F_{34}+F_{31}F_{24}+F_{23}F_{14})\ .
\eea
For even $l$ the $q$-integral vanishes since the integrand is odd under
$q\rightarrow -q$. For odd $l$, one the other hand, the momentum integral
gives in the coincident limit,
\bea
\int \frac{d^2p}{(2\pi)^2} (p^2)^{n-1} e^{-s(p^2)^n}\times\int \frac{dq}{2\pi}C_l q^{l-1}
e^{-sC_l^2(q^2)^l}=\frac{1}{4\pi s}\frac{1}{n}\times \frac{1}{\sqrt{4\pi s}}\frac{{\rm sgn}( C_l)}{l}\ .
\eea
This brings us to
\bea
N_{\vec P}=\left\{\begin{array}{cl}{\rm sgn}( C_l)\cdot n&\qquad \hbox{ for odd }l \\ & \\
0&\qquad \hbox{ for even }l\end{array}\right. \ ,
\eea
where $l$ does not add to the winding number but rather
turn it on or off, depending on its value modulo 2.

\section{Diagrammatic Computation of the Anomaly}

One of the point that remains unresolved in the above anomaly
computation via the Fujikawa viewpoint was the nature of the
Wick rotation. Although the computation stands well-defined
as a generalized index problem for a spinor valued in a
vector bundle over $\IR^d$, its relation to quantum theories
in real world with the Lorentzian signature needs to be clarified
further when the spatial momentum mixes in with the frequency.
The usual Wick rotation $-it\rightarrow \tau$ no longer works,
because of such a mix, so it is necessary to readdress the
issue from Lorentzian viewpoint by computing the usual Feynman
diagram and see how its result is connected to those above.
In particular, this will give us a clearer picture of exactly
what Wick rotation we have effectively performed and
also clarify why the general form of the anomaly remains
intact, modulo a multiplicative constant, despite the higher
inverse power of the momentum in the propagators.

For this, we consider the generalized fermion action
$S=\int_x {\cal L}$ where
\bea
{\cal L}=-\bar\psi \gamma^\mu \cP_\mu(-iD)\psi +iM\bar\psi \psi\,.
\eea
It is a massive fermion of mass $M$, and the true object we need is a subtraction between $M=0$ fermion and the Pauli-Villars fermion of $M$ in $M\to \infty$ limit. Once we have this subtraction, the loop integral is finite and we are free to do shift/change of variable of loop variables. When we do such operations in the following, it is understood that we do the same simultaneously for both $M=0$ fermion and Pauli-Villars fermion contributions in the regularized integrand, so that it is justified.

We assume that $\cP_\mu(-iD)$ is expandable in power series, and we show details for a particular order $n$ term explicitly, and wherever we can replace the order $n$ expression with
the general $\cP_\mu(k)$ we will do so.
The action with an order $n$ term is
\bea
&&-\bar\psi \gamma^\mu \cP_\mu(-iD)\psi +iM\bar\psi \psi \cr\cr
&=&- C_\mu^{\alpha_1\alpha_2\cdots\alpha_n}\bar\psi \gamma^\mu(-iD)_{\alpha_1}(-iD)_{\alpha_2}\cdots (-iD)_{\alpha_n}\psi +iM\bar\psi\psi\,.
\eea
Either by the Noether method or by introducing auxiliary chiral gauge field $A_5$ in $D=\partial+A+A_5 \gamma^5$ and differentiating the action with respect to $A_5$, the chiral current is obtained as
\bea
j^\mu_A=\sum_{s=1}^n C_\nu^{\alpha_1\cdots\alpha_{s-1}\mu \alpha_{s+1}\cdots\alpha_n} ((iD)_{\alpha_{s-1}}\cdots (iD)_{\alpha_1}\bar\psi)\gamma^{\nu}\gamma^5 (-iD)_{\alpha_{s+1}}\cdots (-iD)_{\alpha_{n}}\psi\,.
\eea
Note that $D\psi=(\partial+A)\psi$ and $D\bar\psi=(\partial-A)\bar\psi$.
Although we don't use the conserved vector current $j^\mu$, it has the same form as the above except $\gamma^5$.
Using the classical equation of motion, it is easily seen that
\bea
\partial _\mu j^\mu_A=2M \bar\psi \gamma^5 \psi\,.\label{classical}
\eea

If there was no UV divergence in the correlation functions involving $j^\mu_A$, the (\ref{classical}) would imply a Ward identity for the correlation functions,
\bea
\partial_\mu\langle j^\mu_A\cdots \rangle=2M \langle \bar\psi\gamma^5\psi\cdots\rangle
\eea
which could be seen diagrammatically order by order in the loop expansion.
We have checked this explicitly at 1-loop up to second order in the background gauge fields, that is relevant to the chiral anomaly.
In showing this, one needs to shift or change the loop momenta, which is valid only when the integrand is UV finite. For a divergent diagram, such as the one we need to compute for chiral anomaly, these operations are allowed only after subtracting the regularization contribution to make it finite, in our case, the Pauli-Villars fermion.
The regularized chiral current is therefore
\bea
j^\mu_{A,{\rm reg}}=j^\mu_{A,M=0}-j^\mu_{A,M}\,,
\eea
which has now the valid Ward identity
\bea
\partial_\mu \langle j^\mu_{A,{\rm reg}}\cdots\rangle= -2M\langle \bar\psi_M \gamma^5 \psi_M\cdots\rangle\,,
\eea
where $\psi_M$ is the Pauli-Villars fermion. The right-hand side of the above will be seen to be finite in the 1-loop order that is sufficient to derive our chiral anomaly,\footnote{If the right-hand side remains divergent, although the degree of divergence should be reduced, one needs to introduce the second pair of Pauli-Villars fields to subtract such divergence. One continues to introduce the necessary Pauli-Villars fields until it becomes finite.} and the chiral anomaly, whatever it is, is obtained by
\bea
\cA=\lim_{M\to\infty} -2M\langle \bar\psi_M \gamma^5 \psi_M\rangle\,.\label{M}
\eea

There is another view point that leads to the same formula for chiral anomaly.
Consider a physical massive fermion of mass $M$. Its chiral current has the Ward identity
\bea
\partial_\mu j^\mu_A=2M\bar \psi\gamma^5 \psi+\cA\,.
\eea
As $M\to \infty$, the fermion should decouple from the low energy regime, and the symmetry current and its Ward identity should be provided and saturated by other low energy degrees of freedom in $M\to\infty$ limit. This means that the right-hand side of the above must vanish in $M\to\infty$ limit.

\begin{figure}[t]
 \centering
 \includegraphics[height=5.5cm]{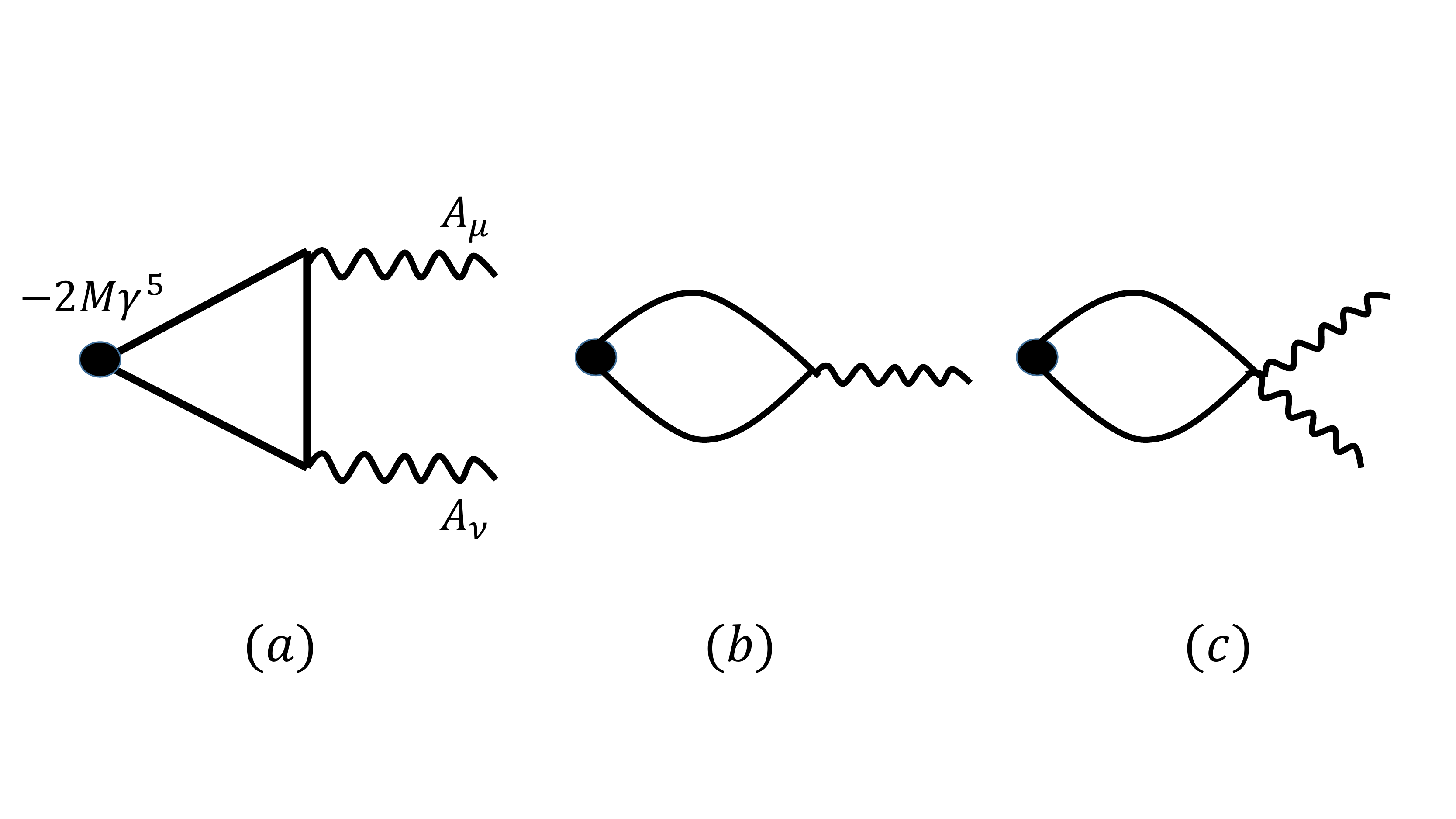}\caption{The Feynman diagrams for $-2M\langle \bar\psi_M \gamma^5 \psi_M\rangle$. \label{fig1}}
 \end{figure}
\begin{figure}[t]
 \centering
 \includegraphics[height=5.5cm]{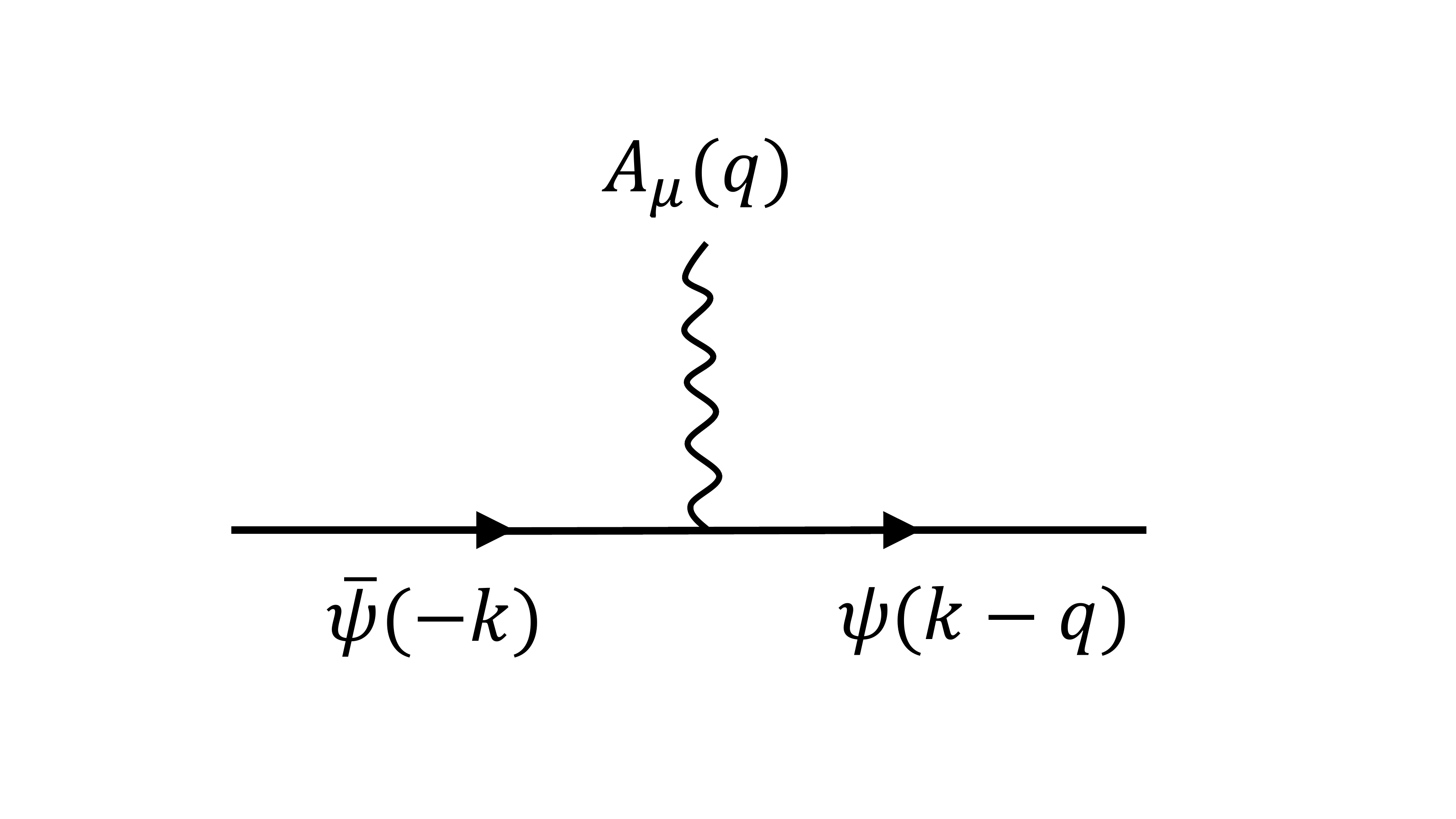}\caption{The Feynman diagram for the interaction vertex. \label{fig2}}
 \end{figure}

In the diagrammatic evaluation of (\ref{M}) up to second order in the background gauge fields, there are three diagrams to compute as shown in Figure \ref{fig1}.
Since a non-vanishing $\gamma^5$ trace requires at least four $\gamma$ matrices, and each vertex and propagator contains at most one $\gamma$ matrix, the diagrams (b) and (c) vanish trivially, and we only consider the triangle diagram (a).
The propagator in momentum space as a spinor matrix is\footnote{We denote the 4-momentum in the Lorentzian signature
by $k$, to be distinguished from the Euclidean 4-momentum $K$ or from the spatial 3-momentum $\vec k$. The precise relation between the Euclidean $K$ we used in Section 2 and Lorentzian $k$ is more subtle than for the usual Dirac fermions. See the last part of this section.}
\bea
\langle \psi(k) \bar\psi(k')\rangle=(2\pi)^4\delta^{(4)}(k+k') {(-i)\over \gamma^\mu \cP_\mu(k)-iM}\,.
\eea
We need the interaction vertex only up to first order in the gauge field,
\bea
iS\sim -\sum_{s=1}^n C_\mu^{\alpha_1\cdots\alpha_n} ((i\partial)_{\alpha_{s-1}}\cdots (i\partial)_{\alpha_1}\bar\psi)\gamma^{\mu} A_{\alpha_s} (-i\partial)_{\alpha_{s+1}}\cdots (-i\partial)_{\alpha_{n}}\psi\,.
\eea
The Feynman rule for this vertex as shown in Figure \ref{fig2} is given by
\bea
-\sum_{s=1}^n C_\mu^{\alpha_1\cdots\alpha_n}  k_{\alpha_{1}}\cdots k_{\alpha_{s-1}}(k-q)_{\alpha_{s+1}}\cdots (k-q)_{\alpha_{n}}\gamma^{\mu} A_{\alpha_s}(q)\equiv \Gamma(k,q)\cdot A(q)\,,
\eea
where $\psi$ carries a momentum $k-q$, $\bar\psi$ a momentum $-k$, and the momentum of the gauge field $A$ is $q$. Note that
\bea
\Gamma(k,0)\cdot A(q)=-\gamma^\mu{\partial \cP_\mu(k)\over\partial k_\nu} A_\nu(q)\,,
\eea
which will be used later.

It is easy to write down the expression of $-2M\langle \bar\psi_M\gamma^5\psi_M\rangle$ of momentum $p$ as (where $\psi$ carries a momentum $k$ and $\bar\psi$ a momentum $p-k$)
\bea
2M\int_q\int_k &{\rm Tr}&\Bigg(\gamma^5 {(-i)\over \gamma\cdot \cP(k)-iM} \Gamma(k,q)\cdot A(q) {(-i)\over \gamma\cdot \cP(k-q)-iM}\nonumber\\ &\times& \Gamma(k-q,p-q)\cdot A(p-q) {(-i)\over \gamma\cdot \cP(k-p)-iM} \Bigg)\,,\nonumber\\
\eea
where $\int_k=\int{d^4k\over (2\pi)^4}$ is the loop integration. This is equal to
\bea
2iM\int_q\int_k  {{\rm Tr} (\cI)\over (\cP(k)^2+M^2-i\epsilon)(\cP(k-q)^2+M^2-i\epsilon)(\cP(k-p)^2+M^2-i\epsilon)}\,,\label{loopint}
\eea
where $\cP^2=\cP_\mu \cP^\mu$, and
\bea
{\rm Tr}(\cI)&=&{\rm Tr}\Bigg(\gamma^5 (\gamma\cdot \cP(k)+iM) \Gamma(k,q)\cdot A(q)(\gamma\cdot \cP(k-q)+iM)\nonumber\\ &\times& \Gamma(k-q,p-q)\cdot A(p-q) (\gamma\cdot \cP(k-p)+iM) \Bigg)\,.
\eea
Note that we introduced $-i\epsilon$ for the time ordered correlation functions.

The non-zero $\gamma^5$ trace in the above requires precisely four $\gamma$ matrices, and
\bea
{\rm Tr}(\cI)&=&iM({\rm Tr}(\cI_1)+{\rm Tr}(\cI_2)+{\rm Tr}(\cI_3))\,,
\eea
where
\bea
{\rm Tr}(\cI_1)&=&{\rm Tr}\Bigg(\gamma^5(\Gamma(k,q)\cdot A(q))( \gamma\cdot \cP(k-q))(\Gamma(k-q,p-q)\cdot A(p-q)) (\gamma\cdot \cP(k-p)) \Bigg)\nonumber\\
{\rm Tr}(\cI_2)&=&{\rm Tr}\Bigg(\gamma^5 (\gamma\cdot \cP(k))( \Gamma(k,q)\cdot A(q)) (\Gamma(k-q,p-q)\cdot A(p-q) )(\gamma\cdot \cP(k-p)) \Bigg)\nonumber\\
{\rm Tr}(\cI_3)&=&{\rm Tr}\Bigg(\gamma^5 (\gamma\cdot \cP(k))( \Gamma(k,q)\cdot A(q))(\gamma\cdot \cP(k-q))( \Gamma(k-q,p-q)\cdot A(p-q)) \Bigg)\,.\nonumber\\
\eea
Since $\gamma^5$ trace is totally anti-symmetric with respect to four $\gamma$ matrices in each $\cI_{1,2,3}$, we can shuffle four factors inside each $\cI_{1,2,3}$ up to sign changes.
By the same reason, we can add any multiple of one factor to the other factor without changing the result. Adding ${\rm Tr}(\cI_2)$ and ${\rm Tr}(\cI_3)$, we have
\bea
{\rm Tr}(\cI_2)+{\rm Tr}(\cI_3)&=&{\rm Tr}\Bigg(\gamma^5 (\gamma\cdot \cP(k))( \Gamma(k,q)\cdot A(q)) (\Gamma(k-q,p-q)\cdot A(p-q) )\cr\cr&&\hskip 3cm \times (\gamma\cdot \cP(k-p)-\gamma\cdot\cP(k-q)) \Bigg)\,.
\eea
Subtracting the second factor from the last factor in ${\rm Tr}(\cI_1)$,
\bea
{\rm Tr}(\cI_1)&=& {\rm Tr}\Bigg(\gamma^5(\Gamma(k,q)\cdot A(q))( \gamma\cdot \cP(k-q))(\Gamma(k-q,p-q)\cdot A(p-q)) \cr\cr&&\hskip 3cm \times (\gamma\cdot \cP(k-p)-\gamma\cdot\cP(k-q)) \Bigg)\,.
\eea
Adding these, we have
\bea
&&{\rm Tr}(\cI_1)+{\rm Tr}(\cI_2)+{\rm Tr}(\cI_3)\cr\cr
&=&{\rm Tr}\Bigg(\gamma^5 (\gamma\cdot \cP(k)-\gamma\cdot\cP(k-q))
 \times (\gamma\cdot \cP(k-p)-\gamma\cdot\cP(k-q)) \cr\cr&&
\hskip 5mm \times ( \Gamma(k,q)\cdot A(q))(\Gamma(k-q,p-q)\cdot A(p-q) ) \Bigg)\,.
\eea

The first factor vanishes linearly in small $q$ limit,
\bea
\gamma\cdot \cP(k)-\gamma\cdot\cP(k-q)\approx \gamma^\mu{\partial \cP_\mu(k)\over\partial k_\nu}q_\nu+\cdots\,,
\eea
which, together with $A(q)$, gives a first order derivative of $A$ in coordinate space.
The same is true for the second factor,
\bea
\gamma\cdot \cP(k-p)-\gamma\cdot\cP(k-q)\approx
 -\gamma^\mu{\partial \cP_\mu(k)\over\partial k_\nu}(p-q)_\nu+\cdots\,,
\eea
that combines with $A(p-q)$ to give another derivative of $A$ in coordinate space.
If we truncate higher order derivatives of $A$ in coordinate space, we need to put $q$
and $(p-q)$ in the other factors to zero.\footnote{We can make it less {\it ad hoc} by the observation that the expansion parameter is either $q/k$ or $(p-q)/k$. In $M\to\infty$ limit, our final result of chiral anomaly, which is finite, is dominated by the integration region of $k\sim M$, and the higher derivative terms are suppressed by additional powers of $\partial/M$, which vanish in $M\to\infty$.}
Then the third and fourth factors become
\bea
\Gamma(k,q)\cdot A(q)\to \Gamma(k,0)\cdot A(q)=-\gamma^\mu{\partial \cP_\mu(k)\over\partial k_\nu} A_\nu(q)\,,
\eea
and
\bea
\Gamma(k-q,p-q)\cdot A(p-q) \to -\gamma^\mu{\partial \cP_\mu(k)\over\partial k_\nu} A_\nu(p-q)\,.
\eea
Computing the trace, we obtain
\bea
{\rm Tr}(\cI)= 4M\, {\rm det}\left({\partial \cP_\mu\over\partial k_\nu}\right)\epsilon^{\alpha\beta\alpha'\beta'}(iq_\alpha)A_\beta(q)(i(p-q)_{\alpha'})A_{\beta'}(p-q)\,,
\eea
and in the remaining loop integral of (\ref{loopint}) in $M\to\infty$ limit, we can neglect $q$ and $p$ in the denominator since it is dominated by $k\sim M \gg p,q$ region, and we arrive at the result for the chiral anomaly
\bea
&&8iM^2 \int_q \epsilon^{\alpha\beta\alpha'\beta'}(iq_\alpha)A_\beta(q)(i(p-q)_{\alpha'})A_{\beta'}(p-q) \cr\cr &&\hskip 2cm \times \int_k {\rm det}\left({\partial \cP_\mu(k)\over\partial k_\nu}\right) {1\over (\cP(k)^2+M^2-i\epsilon)^3}\,,\nonumber\\
\eea
which is in coordinate space,
\bea
\cA=\epsilon^{\alpha\beta\alpha'\beta'}F_{\alpha\beta}F_{\alpha'\beta'}
 \int_k {\rm det}\left({\partial \cP_\mu(k)\over\partial k_\nu}\right) {2iM^2\over (\cP(k)^2+M^2-i\epsilon)^3} \,.
\eea
The above $k$ integration can be done in $\cP$ space up to the winding number $N_\cP$ explained before,
\bea
\int_k {\rm det}\left({\partial \cP_\mu(k)\over\partial k_\nu}\right) {2iM^2\over (\cP(k)^2+M^2-i\epsilon)^3}=N_\cP\int {d^4 \cP\over (2\pi)^4} {2iM^2\over (\cP^2+M^2-i\epsilon)^3}\,,
\eea
and the Wick rotation $\cP^0\to i\cP_E^0$ gives
\bea
\int {d^4 \cP\over (2\pi)^4} {2iM^2\over (\cP^2+M^2-i\epsilon)^3}=
-\int {d^4 \cP_E\over (2\pi)^4} {2M^2\over (\cP_E^2+M^2)^3}=-{1\over 16\pi^2}\,,
\eea
which finally gives the chiral anomaly
\bea
\cA=-N_\cP{1\over 16\pi^2} \epsilon^{\alpha\beta\alpha'\beta'}F_{\alpha\beta}F_{\alpha'\beta'}\,.
\eea
This result is the same as (\ref{Anomaly}) except an overall factor 2,
which has a well-known origin: Here we are computing the anomaly of
a single 4-component Dirac-like fermions while, in Sections 2 and 3,
we computed for 2-component chiral fermions.
Note that, again, the winding number $ N_\cP$ factors out cleanly;
the final form of the integral is expressed in terms of $\cP$, in
place of $K$, such that the only surviving information about $\cP$ is
how many times $\cP$ covers the momentum space $\IR^4$.

Specializing to the case of a non-relativistic Weyl semimetal, where
\bea
\cP_0(k)=k_0+\Delta(\,\vec k\,)\,,\quad \cP_a(k)=P_a(\,\vec k\,)\,,\quad a=1,2,3\,,
\eea
we have
\bea
&&\int_k {\rm det}\left({\partial \cP_\mu(k)\over\partial k_\nu}\right) {2iM^2\over (\cP(k)^2+M^2-i\epsilon)^3} \\ &=&
\int {d^3\vec k\over (2\pi)^3}{\rm det}\left({\partial P_a(\,\vec k\,)\over\partial k_b}\right) \int {dk^0\over (2\pi)}{2iM^2\over (-(k^0-\Delta(\,\vec k\,))^2+\vec P(\,\vec k\,)^2+M^2-i\epsilon)^3}\,.\nonumber
\eea
Note that $\Delta(\,\vec k\,)$ can be simply removed by a shift of $k^0$ integration, and the chiral anomaly is not affected by $\Delta(\,\vec k\,)$. Since $\Delta(\,\vec k\,)$ determines the shape of the Fermi surface by $|\vec P(\,\vec k\,)|=-\Delta(\,\vec k\,)$\footnote{The Hamiltonian for the right-handed Weyl component is $\vec P(\,\vec k\,)\cdot\vec\sigma +\Delta(\,\vec k\,)$.}, the chiral anomaly is independent of occupation number of states, but is determined only by the topology of the level crossing point.

Performing the Wick rotation of $k^0$ after shifting to remove $\Delta(\,\vec k\,)$, the $k_E^0$ integration is easily done to give
\bea
-{3\over 8}M^2\int {d^3\vec k\over (2\pi)^3}{\rm det}\left({\partial P_a(\,\vec k\,)\over\partial k_b}\right){1\over (\vec P(\,\vec k\,)^2+M^2)^{5/2}}\,.
\eea
Again, up to the winding number $N_{\vec P}$ of the map $\vec k\to \vec P(\,\vec k\,)$, the above integral can be done in $\vec P$ space,
\bea
-N_{\vec P}{3\over 8}M^2\int {d^3\vec P\over (2\pi)^3}{1\over (\vec P^2+M^2)^{5/2}}=-N_{\vec p}\,{1\over 16\pi^2}\,,
\eea
which reproduces the previous result by the index theorem.

One important point of this computation is the Wick rotation, $\cP^0\to i\cP_E^0$,
which is not the same as Wick rotation of the Lorentzian time to the Euclidean
time. In the presence of $\Delta(\,\vec k\,)$, the latter cannot be justified
as the rotation of the contour $k_0\rightarrow ik_4$ can encounter a pole
along the way. On the other hand, as far as the diagrammatic computation
goes, the Wick rotation is merely a trick that allows efficient computation.
Therefore, a different contour choice in the complex plane of $k_0$ such that
$\cP^0\to i\cP_E^0$ occurs is perfectly acceptable, as long as it is
consistent with the idea of the Feynman propagator.

\section{Generalized Spinors in Odd Dimensions}

In odd spacetime dimensions, neither the usual $\IZ$-valued
index nor the chiral anomaly exists. Instead one finds
discrete anomalies, which have been the focus of active
investigations recently in the context of the
topological insulators. In all such
investigations, the main physical quantity of interest
is the phase of the partition function. When the fermion
is massless, the reality of the action naively implies
that  the partition function is real, yet actual path integral
generically produces a phase factor. And, much as in the anomaly
computation in even dimensions, this becomes apparent
under the inevitable regularization of the path integral.

This phase has been
computed in Ref.~\cite{AlvarezGaume:1984nf} and shown to
be proportional to the eta-invariant,
\bea\label{etadef}
\eta\equiv \lim_{s\rightarrow 0} \sum {\rm sgn}(\lambda)|\lambda|^{-s}
\eea
with the sum over the eigenvalues of the Dirac operator,
which measures some notion of the asymmetry of the eigenvalues
of the Dirac operators under a sign flip.
In the physics community, this is sometimes misrepresented
by its cousin, namely the Chern-Simons action,
\bea\label{etaCS}
\frac{\pi}{2} \eta\;\biggr\vert_{-i\sigma^a (\partial_a+A_a)} & = & \frac12 S_{\rm CS}(A)+\cdots\ .
\eea
The ellipsis denote certain non-local terms,
which makes the entire expression gauge-invariant
even if the Chern-Simons coefficient is not properly quantized:
this has to be since the eigenvalues that define the eta-invariant
are always gauge invariant no matter what \cite{AlvarezGaume:1984nf}.

In this section, we will extend our investigation using
the modified heat kernel to Dirac fermions in odd spacetime
dimensions, with high-derivative Dirac operators as before.
We will explicitly work out $d=3$ Dirac fermion with the
higher-derivative action
\bea
S_3=-\int d^3x\; \bar \psi \cQ_3\psi\ ,
\eea
for which the relevant Dirac operator is
\bea
\cQ_3=\sigma^aP_a(-i\vec D)\ ,
\eea
again with an arbitrary smooth functions $P_a$.

We are interested in the purely imaginary piece of
\bea
W(A)=-\log\left(\int[D\bar \psi D\psi] \;e^{-S_3}\right)\ ,
\eea
which we will denote as $W^{\rm odd}$. We will find that
this phase, or equivalently, the eta-invariant
of this modified Dirac operator, is such that
\bea\label{etaCS}
W^{\rm odd}=\pm \,i\frac{\pi}{2} \eta\;\biggr\vert_{\sigma^a P_a}
&= &N_{\vec P}\cdot \left(\pm \,\frac{i}{2} S_{\rm CS}(A)+\cdots\right)
\eea
with the same winding number $N_{\vec P}$ of the map $P_a$,
as in (\ref{NvP}). The same notation $P_a$ as in
Section \ref{sec:NR} is used here because eventually we will
connect to $d=4$ system with the 4-th direction treated
as the normal to a $d=3$ boundary. For Majorana
fermions, the discussions below carry over straightforwardly
by multiplying the effective action by $1/2$.

\subsection{Eta-Invariant for Generalized Spinors}

As has been studied thoroughly for $\vec P=-i\vec D$ in Ref.~\cite{AlvarezGaume:1984nf},
and also reviewed more recently in the context of topological insulators in Ref.~\cite{Witten:2015aba},
the path integral of such a fermion would lead to an effective
action whose imaginary part is particularly simple to compute.
We start with the Pauli-Villars regularized
partition function
\bea
\prod\frac{\lambda}{\lambda+iM}
\eea
with eigenvalues $\lambda$ and the regulator mass $M\rightarrow \pm\infty$.
The phase of individual pieces are $- i (\pi/2) {\rm sgn}(M) (\lambda)$,
leading to \cite{AlvarezGaume:1984nf}
\bea
W(A)^{\rm odd} = \pm i\frac{\pi}{2}\eta(A)
\eea
with (\ref{etadef}).

Exactly  the same reasoning goes through
even if we replace the usual first order Dirac operator
by $\cQ_3  = \sigma^aP_a(-i\vec D)$,
\bea
W(A^{(R)})^{\rm odd}_{\vec P} = \pm i\frac{\pi}{2}\eta_{\vec P}(A^{(R)})\ ,
\eea
where we put the subscript $\vec P$ to remember that the
Dirac operator is replaced by $\cQ_3$. The eta-invariant has the usual
definition,
\bea
\eta_{\vec P}(A^{(R)})= \lim_{s\rightarrow 0} \sum {\rm sgn}(\lambda_{\cQ_3})|\lambda_{\cQ_3}|^{-s}\ ,
\eea
where the sum is now over all eigenvalues of the operator $\cQ_3$.
We labeled the gauge field by the representation $R$ of $\psi$ explicitly
for the purpose of clarifying the normalization.

As noted already, the local part of this effective action
for the usual $P_a=-iD_a$, on the other hand, is the well-known
Chern-Simons action,
\bea\label{etaCS}
\frac{\pi}{2} \eta_{\vec P=-i\vec D}(A^{(R)})=  t_2(R) \cdot\frac12 S_{\rm CS}(A) +\cdots\ ,
\eea
where $S_{\rm CS}(A)$ is the properly quantized Chern-Simons
action such that its half appears when we integrate out complex
unit-charge $\psi$ under $U(1)$ or $\psi$ in the fundamental
representation in $SU(N)$ for example. $t_2(R)$ is the quadratic
invariant that keeps track of the gauge representation $R$. When
$R$ is the adjoint representation, for example, it
happens to be equal to the twice the dual Coxeter number,
$t_2(R={\rm adjoint})=2h^\vee$.

Here, we wish to show
that, with general $\vec P$, the above relations is modified simply as
\bea\
\frac{\pi}{2} \eta_{\vec P(-i\vec D)}=  N_{\vec P}\cdot  t_2(R) \cdot\frac12  S_{\rm CS}(A) +\cdots
\eea
with the same winding number $N_{\vec P}$ that appears in $d=4$
chiral anomaly.
Upon introducing a Pauli-Villars regulator with $M$,
again, we write the imaginary part of the effective action as
\bea
W(A^{(R)})^{\rm odd}_{\vec P}  = \frac12\sum \left[
\log(\lambda_{\cQ_3}+iM)-\log(\lambda_{\cQ_3}-iM)
\right]
\biggr\vert^{\rm finite}_{M\rightarrow \pm \infty}\ .
\eea
The way to the relation of type (\ref{etaCS}) comes from
how this effective action varies as we vary the gauge field,
\bea
\delta W(A^{(R)})^{\rm odd}_{\vec P} & = &\frac12\sum \left[
\frac{\delta\lambda_{\cQ_3}}{\lambda_{\cQ_3}+iM}
-\frac{\delta\lambda_{\cQ_3}}{\lambda_{\cQ_3}-iM}
\right]
\biggr\vert_{M\rightarrow \pm \infty} \cr\cr
&=&-iM \cdot {\rm Tr} \left[
\frac{\delta{\cQ_3}}{(\cQ_3)^2+M^2}\right]
\biggr\vert_{M\rightarrow \pm \infty}\cr\cr
&=& -iM \cdot \int_0^\infty ds \;{\rm Tr} \left[
\delta{\cQ_3}e^{-s((\cQ_3)^2+M^2)}\right]
\biggr\vert_{M\rightarrow \pm \infty}\ .
\eea
Although the scaling of $s$
looks different from those we used in the anomaly computation,
the content is no different because of $e^{-sM^2}$ term
in the integrand. The large $M$ limit confines $s$ integral
effectively to a region of $s< 1/M^2$, so again the small
$s$ expansion of the heat kernel becomes sufficient.
And the computation again boils down to a heat kernel
one in the coincident limit.

As such, in $d=3$, the first iteration suffices,
\bea
G^{(1)}_s(y;x) =\int_0^s dt \int d^3z \;G_{s-t}^{(0)}(y;z)
\times ((\cQ_3)^2-\vec P(-i\vec \partial)^2) G_t^{(0)}(z;x)
\eea
with
\bea
G_s^{(0)}(x+X;x)\equiv\langle x+X\vert e^{-s\vec P(-i\vec \partial)^2}\vert x\rangle=\int\frac{d^3k}{(2\pi)^3} e^{i\vec k\cdot \vec X}e^{-s\vec P(\,\vec k\,)^2}\ ,
\eea
as we can see explicitly below.

With nontrivial $\vec P$, we find again in the momentum space
an expansion around a generic point $x$
\bea
\delta \cQ_3 &=& \sigma^a\tilde\partial^b P_a(\,\vec k\,) \delta A_b(x)  +\cdots \cr\cr
(\cQ_3)^2&=&\vec P(\,\vec k\,)^2+i\epsilon^{abc}\sigma^aF_{fg}(x)\tilde\partial^fP_b(\,\vec k\,)\tilde\partial^gP_c(\,\vec k\,)+\cdots\ ,
\eea
where the ellipsis denotes again those terms that are
suppressed by small $s$, or equivalently large $M$, scaling.
Since $\delta \cQ_3$ carries a single $\sigma^a$, the trace
over the spinor requires another $\sigma^a$, which can be
supplied by the field strength term in $\cQ_3^2$ shown above.

The trace over the two-component spinor indices leaves behind,
\bea\label{deltaW}
&&\delta_{ A(x)} W(A^{(R)})^{\rm odd}_{\vec P} \cr\cr
& = &i M \cdot \int_0^\infty ds\;s\, e^{-sM^2}
\int \frac{d^3k}{(2\pi)^3}   e^{-s\vec P(\,\vec k\,)^2}{\rm det}
\left(\frac{\partial P_f(\,\vec k\,)}{\partial k_g}\right){\rm tr}_R\left( \epsilon^{abc}\delta A_a(x) F_{bc}(x) \right)
\biggr\vert_{M\rightarrow \pm \infty}\cr\cr
& = & N_{\vec P}\cdot \left(i M\cdot  \int_0^\infty ds\;s\, e^{-sM^2}
\int \frac{d^3P}{(2\pi)^3}   e^{-s\vec P^2}\cdot
{\rm tr}_R\left( \epsilon^{abc}\delta A_a(x) F_{bc}(x) \right)
\biggr\vert_{M\rightarrow \pm \infty}\right)\ ,
\eea
at a generic point $x$, $\vec k$-integrals and an $s$-integral,
and $N_{\vec P}$ is the same winding number we encountered
in Section \ref{sec:NR}.

Let us count the factor of $M$ to ensure one ends up with
a finite quantity: the three $P$ integrations will generate
$s^{-3/2}$, so the final $s$ integral will be of the form
\bea
M\cdot \int_0^\infty ds \;s^{-1/2}e^{-sM^2} =
{\rm sgn}(M)\cdot \int_0^\infty d\tilde s \;\tilde s^{-1/2}e^{-\tilde s} =
{\rm sgn}(M)\cdot\sqrt{\pi}\ .
\eea
$M$ is scaled out leaving behind only its sign, and
as in the previous anomaly computation,  other terms
in $(\cQ_3)^2$ can at most contribute pieces that
scale inversely with the large $M$.

This way, the quantity inside
the large parenthesis remains finite in the limit and
produces the variation of the imaginary part of
the effective action due to a single two-component
spinor with $\vec P=-i\vec D$. For $\vec P=-i\vec D$,
in fact, this is precisely how one shows the relation
(\ref{etaCS}) by starting with the above and integrating
over $\delta\vec A(x)$ back to the Chern-Simons action.
Therefore, we obtain at the end of the computation
\bea
W(A^{(R)})^{\rm odd}_{\vec P} = N_{\vec P}\cdot \left(\pm i t_2(R) \cdot\frac12  S_{\rm CS}(A) +\cdots \right)\ ,
\eea
or,
\bea
W(A^{(R)})^{\rm odd}_{\vec P} =N_{\vec P}
\cdot \left(
\pm i \frac{\pi}{2} \eta_{\vec P=-i\vec D}(A^{(R)})\right) = N_{\vec P}\cdot W(A^{(R)})^{\rm odd}_{\vec P= -i\vec D}\ ,
\eea
given the general relation
between the Chern-Simons action and the eta-invariant
in (\ref{etaCS}), and  also from how the effective action
has to be gauge invariant.

Generalization to higher dimension $d=2n-1$ is also straightforward.
It is  clear that the variation of the effective
action (\ref{deltaW}) can be easily extended to
$d=(2n-1)$ and produce,
\bea
\delta_{A(x)} W^{\rm odd}_{\cQ_{2n-1}} \;\;\sim\;\; N_{\vec P} \cdot {\rm tr}\left(
\delta A\wedge F^{n-1}\right)\ ,
\eea
now expressed as the $d$-form,
for the generalized Dirac operator
\bea
\cQ_{2n-1} =\gamma^a P_a(-\vec D)
\eea
with $2n-1$ dimensional Dirac matrices $\gamma^a$'s.
On the other hand, the Chern-Simons density is defined via
\bea
dS_{\rm CS} = \frac{1}{n!(2\pi i)^n}\int {\rm tr} F^n
\eea
so its variation is such that
\bea
d\left[\delta_{A(x)} S_{\rm CS}\right] &= & \frac{1}{(n-1)!(2\pi i)^n}   \cdot {\rm tr}
\left((d(\delta A)+A\delta A+\delta A A) F^{n-1}\right)\cr\cr
&=& d\left[ \frac{1}{(n-1)!(2\pi i)^n} \cdot {\rm tr} \left((\delta A(x)\wedge F^{n-1}(x)\right)\right]
\eea
which brings us back to
\bea
W(A^{(R)})^{\rm odd}_{\vec P} =N_{\vec P}\cdot \left(\pm  \frac{i}{2}  S_{\rm CS}(A^{(R)}) +\cdots \right)=N_{\vec P}
\cdot \left(
\pm i \frac{\pi}{2} \eta_{\vec P=-i\vec D}(A^{(R)})\right)
\eea
following the same pattern we saw for $d=3$.

\subsection{Boundary Fermions and APS-like Index Theorem}

When $d$ dimensional manifold
$\cM_d$ has a boundary $\Sigma_{d-1}$, an Atiyah-Patodi-Singer (APS)
index problem on $\cM_d$ can be formulated via the extension
of the manifold by attaching a semi-infinite cylinder with
the cross section $\Sigma_{d-1}$. Imposing the square normalizability
condition for the ground states, one finds
\bea
\cI_{\cM_d}^{\rm APS}=\cI_{\cM_d}^{\rm bulk}-\frac{\eta_{\Sigma_{d-1}}}{2}\ ,
\eea
where $\eta_{\Sigma_{d-1}}$ is the usual eta-invariant of $\Sigma_{d-1}$.

Is there a way to extend this result to the generalized Dirac
problem of our kind? Extending the covariant derivative to include
the spin connection, say, $\nabla_\mu$, is no big deal. However,
one also needs a covariantly constant tensor $C_a^{\mu_1\cdots\mu_l}$
such that operators such as
\bea
\gamma^a C_a^{\mu_1\cdots\mu_l}\nabla_{\mu_1}\cdots \nabla_{\mu_l}
\eea
can be used in place of the ordinary $ \gamma^a e_a^\mu\nabla_\mu$
with the vielbein $e_a^\mu$. Such a tensor $C$ implies reduced
holonomy, yet the latter is classified completely and known to be
rather sparse. As such, general higher-derivative Dirac operator
with the curved geometry is generally difficult to construct.

When both $\cM_d$ and $\Sigma_{d-1}$ are flat, one other hand, our
discussions so far does imply an APS-like index theorem.
The easiest is ${\cM_d}=\IT^{d-1}\times \IR_+$ with the boundary
$\Sigma_{d-1}=\IT^{d-1}$ at the origin of $\IR_+$. With the generalized
Dirac operator as in (\ref{nonR})
\bea
\cQ_{d-1}- D_d=\gamma^a P_a (-i\vec D) - D_d\ ,
\eea
and thus
\bea
\cQ_d= \gamma^aP_a(-i\vec D) +\gamma^d (-iD_d) =
\left(\begin{array}{cc}0 & \cQ_{d-1}- D_d \\ \cQ_{d-1}+ D_d & 0\end{array}\right)\ ,
\eea
as in (\ref{nonR4}), with the direction $d$ considered as the
normal to $\Sigma_{d-1}$. We have already seen that the bulk part
of the index is
\bea
\cI_{\cQ_d}^{\rm bulk} = N_\cP\cdot \cI^{\rm bulk}\ ,
\eea
while the eta-invariant is similarly enhanced as
\bea
\frac{\eta_{\cQ_{d-1}}}{2} = N_\cP\cdot \frac{\eta}{2}\ .
\eea
With these, we can easily retrace the steps found in the
appendix of Ref.~\cite{AlvarezGaume:1984nf} with the modified
heat kernel we have discussed so far, and
arrive at a generalized APS index theorem,
\bea
\cI_{\cQ_d}=N_{\vec P}\cdot \left( \frac{1}{n!(2\pi i)^n} \int_{\IT^{2n-1}\times \IR_+} {\rm tr}\left(F^{(R)}\wedge\cdots\wedge F^{(R)}\right)
-\frac{\eta^{\IT^{2n-1}}(A^{(R)})}{2}\right)
\eea
for any even dimensions $d=2n$.

This form of the APS-like index theorem would suffice for
understanding interface between topological insulators and
ordinary insulator, by considering physics very near the
boundary. As before, the factor $N_{\vec P}$
can be realized either as $N_{\vec P}$ many flavors of
ordinary boundary fermions, or a single fermions with
higher order Dirac operator with the winding number $N_{\vec P}$.

\vskip 1cm \centerline{\large \bf Acknowledgment} \vskip 0.5cm

We are grateful to Gil-Young Cho, Paolo Glorioso, Byungmin Kang, Dam Son, Misha Stephanov, and Raju Venugopalan
for helpful discussions. H.-U.Y. appreciates the support and hospitality of KIAS during his visit
when this work was initiated. The work of H.-U.Y. is supported by the U.S. Department of Energy,
Office of Science, Office of Nuclear Physics, with the grant No. DE-FG0201ER41195, and
within the framework of the Beam Energy Scan Theory (BEST) Topical Collaboration.

\end{document}